\documentstyle[amssymb,preprint,aps]{revtex}

\begin{document}
\draft
\title{{\bf General Quantum Resonances of the Kicked Particle}}
\author{{\bf Itzhack Dana}$^{1}${\bf \ and Dmitry L. Dorofeev}$^{1,2}$}
\address{$^{1}$Minerva Center and Department of Physics, Bar-Ilan University,
Ramat-Gan 52900, Israel}
\address{$^{2}$Department of Physics, Voronezh State University, Voronezh 394693,
Russia}
\maketitle

\begin{abstract}
The quantum resonances (QRs) of the kicked particle are studied in a most
general framework by also considering {\em arbitrary} periodic kicking
potentials. It is shown that QR can arise, in general, for {\em any rational}
value of the Bloch quasimomentum. This is illustrated in the case of the
main QRs for arbitrary potentials. In this case, which is shown to be
precisely described by the linear kicked rotor, exact formulas are derived
for the diffusion coefficients determining the asymptotic evolution of the
average kinetic energy of either an incoherent mixture of plane waves or a
general wave packet. The momentum probability distribution is exactly
calculated and studied for a two-harmonic potential. It clearly exhibits new
resonant values of the quasimomentum and it is robust under small deviations
from QR.\newline
\end{abstract}


\begin{center}
{\bf I. INTRODUCTION}
\end{center}

There has been recently a considerable experimental \cite{aoeqr} and
theoretical \cite{kp1,kp2} interest in new remarkable phenomena associated
with the quantum resonances of the periodically kicked particle (KP), either
in the presence or in the absence of gravity. In this paper, we report about
new general results concerning the quantum resonances of the KP in the
absence of gravity. These results extend some of previous work \cite{kp1,kp2}
in a significant way, shedding light on new basic aspects of the problem and
connecting an important case of the system with a well-known model. We start
with a brief summary of previous results \cite{kp1,kp2}, using notation in
Ref. \cite{kp2}. The quantum KP is described by the Hamiltonian 
\begin{equation}
\hat{H}=\frac{\hat{p}^{2}}{2}+kV(\hat{x})\sum_{t}\delta (t^{\prime }-t\tau ),
\label{H}
\end{equation}
where $(x,\ p)$ are the position and momentum of the particle, $k$ is a
nonintegrability parameter, $V(x)$ is a periodic potential, $t$ takes all
the integer values, $t^{\prime }$ is the continuous time, and $\tau $ is the
kicking period. The units are chosen so that the particle mass is $1$, the
Planck's constant $\hbar =1$, and the period of $V(x)$ is $2\pi $. In
practice, $V(x)$ has been always chosen as the standard potential $V(x)=\cos
(x)$. The one-period evolution operator for (\ref{H}), from $t^{\prime }=t+0$
to $t^{\prime }=t+\tau +0$, is given by 
\begin{equation}
\hat{U}=\exp \left[ -ikV(\hat{x})\right] \exp \left( -i\tau \hat{p}
^{2}/2\right) .  \label{U}
\end{equation}
The translational invariance of (\ref{U}) in $\hat{x}$ implies the
conservation of a quasimomentum $\beta $ ($0\leq \beta <1$): The application
of $\hat{U}$ on a Bloch function $\Psi _{\beta }(x)=\exp (i\beta x)\psi
_{\beta }(x)$, where $\psi _{\beta }(x+2\pi )=$ $\psi _{\beta }(x)$, results
in a Bloch function $\Psi _{\beta }^{\prime }(x)=\exp (i\beta x)\psi _{\beta
}^{\prime }(x)$ associated with the same value of $\beta $. Here $\psi
_{\beta }^{\prime }(x)$ is the $2\pi $-periodic function $\psi _{\beta
}^{\prime }(x)=\hat{U}_{\beta }\psi _{\beta }(x)$, where 
\begin{equation}
\hat{U}_{\beta }=\exp \left[ -ikV(\hat{x})\right] \exp \left[ -i\tau \left( 
\hat{p}+\beta \right) ^{2}/2\right] .  \label{Ub}
\end{equation}
The restriction of the operator (\ref{Ub}) to $2\pi $-periodic functions 
$\psi _{\beta }(x)$ allows one to interpret $x$ as an angle $\theta $ and
$\hat{p}$ as an angular-momentum operator $\hat{N}=-id/d\theta $ with integer
eigenvalues $n$. One can then view (\ref{Ub}) as the one-period evolution
operator for a ``$\beta $-kicked rotor'' ($\beta $-KR). Now, an arbitrary KP
wave packet $\Psi (x)$ can be always expressed as a superposition of Bloch
functions, $\Psi (x)=\int_{0}^{1}d\beta \exp (i\beta x)\psi _{\beta }(x)$,
where 
\begin{equation}
\psi _{\beta }(x)=\frac{1}{\sqrt{2\pi }}\sum_{n}\widetilde{\Psi }(n+\beta
)\exp (inx),  \label{phb}
\end{equation}
$\widetilde{\Psi }(p)$ being the momentum representation of $\Psi (x)$. One
then gets the basic relation 
\begin{equation}
\hat{U}^{t}\Psi (x)=\int_{0}^{1}d\beta \exp (i\beta x)\hat{U}_{\beta
}^{t}\psi _{\beta }(x)  \label{KPR}
\end{equation}
for integer ``time'' $t$, connecting the quantum dynamics of the KP with
that of $\beta $-KRs.\newline

For typical irrational values of $\tau /(2\pi )$, $\beta $-KRs are expected
to feature dynamical localization in the angular momentum $n$ \cite{dl,qr},
implying a similar localization of the KP wave packet $\widetilde{\Psi }(p)$
in the momentum $p=n+\beta $ ($n$ and $\beta $ are, respectively, the
integer and fractional parts of $p$). If $\tau /(2\pi )$ is rational, the
usual ($\beta =0$) KR exhibits quantum resonance (QR), i.e., a {\em ballistic} 
(quadratic in time) growth of its kinetic energy \cite{qr}. QR in general 
$\beta $-KRs appears to have been studied only in the case of integer $\tau
/(2\pi )$ (``main'' QRs) with $V(x)=\cos (x)$ \cite{kp1,kp2}. It was found 
\cite{kp2} that QR arises in this case {\em only} for special ``resonant''
values of $\beta $, {\em finite} in number. Thus, QR is exhibited by the KP
only if $\widetilde{\Psi }(p)$ is delta localized on the discrete set of
momenta $p=n+\beta $, with $\beta $ in the finite resonant set. However, QR
leaves a clear fingerprint in the evolution of either a general KP wave
packet (\ref{KPR}) or\ an incoherent mixture of plane waves: In both cases,
which involve {\em all} values of $\beta $, the average kinetic energy grows 
{\em diffusively} (linearly) in time \cite{kp1,kp2}. This diffusive behavior
is robust under small deviations $\epsilon $ of $\tau /(2\pi )$ from
integers, in the sense that it is still observed on time scales $t\propto
|k\epsilon |^{-1/2}$ \cite{kp2}.\newline

In this paper, the results above concerning the QRs of the system (\ref{H})
are extended by also considering {\em arbitrary} periodic potentials $V(x)$.
In Sec. II, we show that the general condition for QR in $\beta $-KRs is
just the {\em rationality} of {\em both} $\tau /(2\pi )$ {\em and} $\beta $.
Thus, a resonant value of $\beta $ may be a {\em general rational} number in 
$[0,\ 1)$ and one then has a (countable) infinity of such values. This is
illustrated in Sec. III in the case of the main QRs for arbitrary $V(x)$. We
show that this case is precisely described by the well-known linear kicked
rotor (linear KR) \cite{lkr1,lkr2}. We focus on this case also in Secs. IV
and V. In Sec. IV, we derive exact formulas for the diffusion coefficients
determining the asymptotic evolution of the average kinetic energy of either
an incoherent mixture of plane waves or a general KP wave packet. For a
mixture uniformly distributed in $\beta $, the formula essentially coincides
with one given by Berry in the context of the linear KR \cite{lkr2}.
Multi-harmonic potentials cause the diffusion coefficient for a general wave
packet to be explicitly dependent on quantum correlations reflecting the
localization features of the initial wave packet in momentum space. In Sec.
V, the momentum probability distribution is exactly calculated and studied
for a two-harmonic potential. It clearly exhibits new resonant values of 
$\beta $ due to the second harmonics and it is found numerically to be robust
under small deviations of $\tau /(2\pi )$ from integers. A summary and
conclusions are presented in Sec. VI.

\begin{center}
{\bf II. GENERAL QR CONDITIONS}
\end{center}

The basic origin of QR is a band quasienergy (QE) spectrum due to some
translational invariance in phase space. For KR systems, this is the
invariance of the evolution operator under translations $\hat{T}_{q}=\exp
(-iq\hat{\theta})$ by $q$ in the angular momentum $\hat{N}$ \cite{qr,cs};
here $q$ must be an integer since $\theta $ is an angle ($0\leq \theta <2\pi 
$). The translational invariance of $\beta $-KRs is expressed by $[\hat{U}
_{\beta },\ \hat{T}_{q}]=0$, where $\hat{U}_{\beta }$ is given by Eq. (\ref
{Ub}) ($\hat{x}\rightarrow \hat{\theta}$, $\hat{p}\rightarrow \hat{N}$).
Using the fact that $\hat{N}$ has integer eigenvalues, one easily finds that 
$[\hat{U}_{\beta },\ \hat{T}_{q}]=0$ is satisfied only if 
\begin{equation}
\frac{\tau }{2\pi }=\frac{l}{q},  \label{tau}
\end{equation}
\begin{equation}
\beta =\frac{r}{l}-\frac{q}{2}\ {\rm mod}(1),  \label{br}
\end{equation}
where $l$ and $r$ are integers. Eq. (\ref{tau}) is the rationality condition
for $\tau /(2\pi )$ while Eq. (\ref{br}) is a formula for the general
resonant values of $\beta $. For definiteness and without loss of
generality, we assume that $l$ and $q$ are positive. Let us now write 
$l=gl_{0}$ and $q=gq_{0}$, where $l_{0}$ and $q_{0}$ are coprime positive
integers and $g$ is the greatest common factor of $(l,\ q)$; the value of
$\tau /(2\pi )=l_{0}/q_{0}$ will be kept fixed in what follows. It is then
clear that $\beta $ in Eq. (\ref{br}) can take {\em any rational} value 
$\beta _{{\rm r}}$ in $[0,\ 1)$ since $g$ can be always chosen so that
$r=(\beta _{{\rm r}}+gq_{0}/2)gl_{0}$ is integer. For given $\beta =\beta _
{{\rm r}}$, we shall choose $g$ as the {\em smallest} positive integer
satisfying the latter requirement, so as to yield the minimal values of 
$l=gl_{0}$ and $q=gq_{0}$. In general, $g>1$, so that $(l,\ q$) are {\em not}
coprime. For the usual KR ($\beta =0$), $g=1$ if $l_{0}q_{0}$ is even and
$g=2$ if $l_{0}q_{0}$ is odd (compare with Ref. \cite{cs}). We denote $\beta
_{{\rm r}}$ by $\beta _{r,g}$, where the integer $r=(\beta _{{\rm r}}+
gq_{0}/2)gl_{0}$ labels all the different values of $\beta _{{\rm r}}$ for
given minimal $g$.\newline

The QE states $\varphi $ for $\beta =\beta _{r,g}$ can be chosen as
simultaneous eigenstates of $\hat{U}_{\beta }$ and $\hat{T}_{q}$: $\hat{U}
_{\beta }\varphi =\exp (-i\omega )\varphi $, $\hat{T}_{q}\varphi =\exp
(-iq\alpha )\varphi $, where $\omega $ is the QE and $\alpha $ is a
``quasiangle'', varying in the ``Brillouin zone'' (BZ) $0\leq \alpha <2\pi
/q $. One may view the Bloch function $\exp (i\beta x)\varphi (x)$ as a
state on the ``quantum torus'' $0\leq x<2\pi $, $0\leq p<q$, with toral
boundary conditions \cite{d} specified by $(\alpha ,\ \beta )$. Using
standard methods \cite{qr,cs}, it is easy to show from the eigenvalue
equations above that at fixed $\alpha $ one has precisely $q$ QE eigenvalues 
$\omega _{b}(\alpha ,\ \beta )$, $b=0,\dots ,\ q-1$. Since $q=gq_{0}$ is
minimal, the BZ is maximal for the given value of $\beta =\beta _{r,g}$.
Then, as $\alpha $ is varied continuously in the BZ, the $q$ eigenvalues
form $q$ QE bands. These bands are expected to be, typically, not all flat 
(with zero width); QR can then arise and $\beta =\beta _{r,g}$ is indeed a
resonant value. In the nontypical case that all the bands are flat, 
$\beta =\beta_{r,g}$ is nonresonant: QR is replaced by a bounded quantum 
motion, the ``quantum antiresonance'' \cite{qar}.

\begin{center}
{\bf III. CASE OF MAIN QRs: CONNECTION WITH THE LINEAR KR}
\end{center}

Previous studies \cite{kp1,kp2} have focused on the important case of the
main QRs ($\tau =2\pi l_{0}$, $q_{0}=1$), assuming that $V(\theta )=\cos
(\theta )$. It was found \cite{kp2} that only $l_{0}$ values of $\beta $ are
resonant (exhibit QR). They are given by Eq. (\ref{br}) with $r=0,\ 1,\dots
,\ l_{0}-1$ and $g=1$ (i.e., $l=l_{0}$ and $q=1$). In this section, we study
the main QRs for arbitrary $V(\theta )$ and we show that all rational values
of $\beta $ in $[0,\ 1)$ are resonant if $V(\theta )$ contains all the
harmonics. The case of $\tau =2\pi l_{0}$ is the only one in which the term 
$(\hat{N}+\beta )^{2}$ in Eq. (\ref{Ub}) ($\hat{x}\rightarrow \hat{\theta}$,
$\hat{p}\rightarrow \hat{N}$) can be replaced by the operator $\hat{N}+2\beta 
\hat{N}+\beta ^{2}$, {\em linear} in $\hat{N}$; this is because $\exp (-i\pi
l_{0}n^{2})=\exp (-i\pi l_{0}n)$ for the integer eigenvalues $n$\ of $\hat{N}
$. Then, after omitting the nonrelevant constant phase factor $\exp (-i\pi
l_{0}\beta ^{2})$, one can express (\ref{Ub}) as follows: 
\begin{equation}
\hat{U}_{\beta }=\exp \left[ -ikV(\hat{\theta})\right] \exp \left( -i\tau
_{\beta }\hat{N}\right) ,  \label{Ubm}
\end{equation}
where $\tau _{\beta }=\pi l_{0}(2\beta +1)$. We identify $\hat{U}_{\beta }$
in Eq. (\ref{Ubm}) as the one-period evolution operator for the well-known
linear KR \cite{lkr1,lkr2} with Hamiltonian $\hat{H}=\tau _{\beta }\hat{N}+
kV(\hat{\theta})\sum_{t=-\infty }^{\infty }\delta (t^{\prime }-t)$; the
corresponding Schr\"{o}dinger equation is exactly solvable for arbitrary
potential $V(\theta )$, 
\begin{equation}
V(\theta )=\sum_{m}V_{m}\exp (-im\theta ).  \label{V}
\end{equation}
Assuming the quantum state of the linear KR to be initially (at $t=0$) an
angular-momentum state, $\psi _{\beta ,0}(\theta )=\exp (in_{0}\theta )/
\sqrt{2\pi }$, the expectation value of the kinetic energy at time $t$ is
given by the exact expression \cite{lkr2,note}: 
\begin{equation}
E_{n_{0},\beta }(t)=\frac{1}{2}\left\langle \psi _{\beta ,t}|\hat{N}
^{2}|\psi _{\beta ,t}\right\rangle =\frac{n_{0}^{2}}{2}+k^{2}
\sum_{m>0}m^{2}|V_{m}|^{2}\frac{\sin ^{2}(m\tau _{\beta }t/2)}{\sin
^{2}(m\tau _{\beta }/2)}.  \label{Ebt}
\end{equation}
If $\beta $ is a typical irrational number, so that also $\tau _{\beta
}/(2\pi )=l_{0}(\beta +1/2)$ is such, the QE spectrum of the linear KR is
pure point \cite{lkr1} and $E_{n_{0},\beta }(t)$ in Eq. (\ref{Ebt}) is a
bounded, quasiperiodic function of time \cite{lkr2}. Consider now a general
rational value of $\beta =\beta _{r,g}$, given by Eq. (\ref{br}) with
$l=gl_{0}$, $q=gq_{0}=g$, and $g$ minimal (see Sec. II). It is easy to show
that $\beta _{r,g}$ can be written, up to a fixed integer shift in $r$ by
$gl_{0}/2$ if both $g$ and $gl_{0}/2$ are even, as $\beta
_{r,g}=r/(gl_{0})-1/2$ ${\rm mod}(1)$, where $r$ and $g$ are coprime; the
corresponding value of $\tau _{\beta }/(2\pi )$ is $r/g$, up to some
additive integer. Then, from work \cite{lkr1}, the QE spectrum consists of $g
$ bands which are all nonflat (i.e., $\beta =\beta _{r,g}$ is resonant) if
there exists at least one integer $j\neq 0$ such that the Fourier
coefficient $V_{jg}$ in Eq. (\ref{V}) is nonzero; otherwise, all the $g$
bands are flat ($\beta =\beta _{r,g}$ is nonresonant, with QR replaced by
quantum antiresonance \cite{qar}). In fact, if $V_{jg}\neq 0$ for some
$j\neq 0$, one finds from Eq. (\ref{Ebt}) (with $\tau _{\beta }=2\pi r/g$) a
ballistic behavior for large $t$, $E_{n_{0},\beta }(t)\approx St^{2}/2$,
where $S=k^{2}g^{2}\sum_{j}j^{2}|V_{jg}|^{2}$ \cite{lkr2}. Thus, if
$V(\theta )$ contains all the harmonics ($V_{m}\neq 0$ for all $m$), all
rational values $\beta _{r,g}$ of $\beta $ are resonant.\newline

As a simple example, let $V(\theta )=\cos (\theta )+\gamma \cos (2\theta )$.
The only nonzero coefficients $V_{m}$ are $V_{\pm 1}=1/2$ and $V_{\pm
2}=\gamma /2$, so that QR arises only for $g=1,\ 2$. The resonant $\beta $
values are $\beta _{r,g}=r/(gl_{0})-1/2$ ${\rm mod}(1)$, where $r=0,\
1,\dots ,\ l_{0}-1$ for $g=1$ and $r=1,\ 3,\dots ,\ 2l_{0}-1$ for $g=2$. The
values for $g=1$ are just the known ones for $V(\theta )=\cos (\theta )$ 
\cite{kp2} while those for $g=2$ are new ones, due entirely to the second
harmonics. See, however, note \cite{note1}.

\begin{center}
{\bf IV. ASYMPTOTIC DIFFUSION OF AVERAGE KINETIC ENERGY}

{\bf A. Incoherent Mixture of Plane Waves}
\end{center}

Let us assume, as in experimental situations \cite{aoeqr}, that the initial
KP state is an incoherent mixture of plane waves $\exp (ipx)$ with momentum
distribution $f(p)$ sufficiently localized in $p$. By decomposing $p$ into
its integer and fractional parts, $p=n+\beta $, the average kinetic energy
of this mixture at time $t$ can be expressed as $\bar{E}(t)=\int_{0}^{1}d
\beta \sum_{n}f(n+\beta )E_{n,\beta }^{\prime }(t)$; here $E_{n,\beta
}^{\prime }(t)$ is the expectation value of the kinetic energy in the state 
evolving from a plane wave. In the case of $\tau =2\pi l_{0}$, on which we 
shall focus, $E_{n,\beta }^{\prime }(t)$ is given by the right-hand side of Eq. 
(\ref{Ebt}) with $n_{0}$ replaced by $p=n+\beta $; this can be easily seen 
from Eqs. (18)-(21) in Ref. \cite{lkr2}. As in Ref. \cite{kp2}, we define 
$f_{0}(\beta )=\sum_{n}f(n+\beta )$ and use the asymptotic (large $t$) relation 
\[
\int_{0}^{1}d\beta f_{0}(\beta )\frac{\sin ^{2}\left[ \pi ml_{0}\left( \beta
+1/2\right) t\right] }{\sin ^{2}\left[ \pi ml_{0}\left( \beta +1/2\right)
\right] }\sim \frac{t}{|m|l_{0}}\sum_{r=0}^{|m|l_{0}-1}f_{0}\left( \beta
_{r,m}\right) ,
\]
where $\beta _{r,m}=r/(|m|l_{0})-1/2$ ${\rm mod}(1)$. We then find that
$\bar{E}(t)$ behaves diffusively for large $t$, $\bar{E}(t)\sim D_{0}t$,
where the diffusion coefficient $D_{0}$ is given by 
\begin{equation}
D_{0}=\frac{k^{2}}{l_{0}}\sum_{m>0}m|V_{m}|^{2}\sum_{r=0}^{ml_{0}-1}f_{0}
\left( \beta _{r,m}\right) .  \label{DIM}
\end{equation}
The simple mixture with $f(n+\beta )=\delta _{n,n_{0}}$ is uniformly
distributed in $\beta $, $f_{0}(\beta )=1$. In this case, $\bar{E}
(t)=\int_{0}^{1}d\beta E_{n_{0},\beta }^{\prime }(t)$, leading to an exact
equality for all times $t$: $\bar{E}(t)=\bar{E}(0)+D_{0}t$ with
$D_{0}=k^{2}/(4\pi )\int_{0}^{2\pi }[dV(\theta )/d\theta ]^{2}d\theta$. The
latter result was essentially obtained by Berry \cite{lkr2} as the kinetic
energy of a linear KR whose value of $\tau $ is completely unknown.\newpage 

\begin{center}
{\bf B. General Wave Packet}
\end{center}

Next, we consider a general KP wave packet (\ref{KPR}), a ``coherent
mixture'' of Bloch waves exhibiting all values of $\beta $. The expectation
value of the kinetic energy in $\Psi _{t}(x)=\hat{U}^{t}\Psi (x)$ is 
\begin{equation}
\left\langle E\right\rangle _{t}=\frac{1}{2}\left\langle \Psi _{t}\left| 
\hat{p}^{2}\right| \Psi _{t}\right\rangle \sim \frac{1}{2}\int_{0}^{1}d\beta
\int_{0}^{2\pi }d\theta \left| \frac{d\psi _{\beta ,t}(\theta )}{d\theta }
\right| ^{2},  \label{ake}
\end{equation}
where the last relation, with $\psi _{\beta ,t}(\theta )\equiv \hat{U}_
{\beta }^{t}\psi _{\beta }(\theta )$, holds for large $t$ provided that
$\left\langle E\right\rangle _{t}$ is unbounded as $t\rightarrow \infty $ 
\cite{kp1}. In fact, we now show that $\left\langle E\right\rangle _{t}$
exhibits an asymptotic diffusive behavior for $\tau =2\pi l_{0}$. In this
case, we easily obtain from Rels. (\ref{Ubm}) and (\ref{V}) that 
\begin{equation}
\psi _{\beta ,t}(\theta )\equiv \hat{U}_{\beta }^{t}\psi _{\beta }(\theta )=
\exp \left[ -ik\bar{V}_{\beta ,t}(\theta )\right]
\psi _{\beta }(\theta -t\tau _{\beta }),  \label{pbt}
\end{equation}
where 
\begin{equation}
\bar{V}_{\beta ,t}(\theta )=\sum_{s=0}^{t-1}V\left( \theta -s\tau _{\beta
}\right) =\sum_{m}V_{m}\frac{\sin (m\tau _{\beta }t/2)}{\sin (m\tau _{\beta
}/2)}e^{-im[\theta -(t-1)\tau _{\beta }/2]}.  \label{Vbt}
\end{equation}
As shown in Ref. \cite{kp1} (Appendix A) for $V\left( \theta \right) =\cos
(\theta )$, with straightforward extension to arbitrary $V\left( \theta
\right) $, the expression (\ref{pbt}) implies that for large $t$ the
dominant contribution of $|d\psi _{\beta ,t}(\theta )/d\theta |^{2}$ to
$\left\langle E\right\rangle _{t}$ in Eq. (\ref{ake}) is $k^{2}\left| \psi
_{\beta }(\theta -t\tau _{\beta })d\bar{V}_{\beta ,t}(\theta )/d\theta
\right| ^{2}$. This contribution can be calculated using Eq. (\ref{Vbt}) and
a relation following from Eq. (\ref{phb}): 
\begin{equation}
\left| \psi _{\beta }(\theta )\right| ^{2}=\frac{1}{2\pi }\sum_{m}C_{\beta
}(m)\exp (im\theta ),  \label{pbc}
\end{equation}
where $C_{\beta }(m)$ are correlations in momentum space, 
\begin{equation}
C_{\beta }(m)=\sum_{n}\widetilde{\Psi }(m+n+\beta )\widetilde{\Psi }^{\ast
}(n+\beta ).  \label{cbm}
\end{equation}
From Eqs. (\ref{Vbt}) and (\ref{pbc}), one can write the Fourier expansion
of $k^{2}\left| \psi _{\beta }(\theta -t\tau _{\beta })d\bar{V}_{\beta
,t}(\theta )/d\theta \right| ^{2}$. After inserting this expansion in Eq.
(\ref{ake}), we get 
\begin{equation}
\left\langle E\right\rangle _{t}\sim \frac{k^{2}}{2}\sum_{m,m^{\prime }\neq
0}mm^{\prime }V_{m}V_{m^{\prime }}^{\ast }B(m,\ m^{\prime };\ t),
\label{ake1}
\end{equation}
where 
\begin{equation}
B(m,\ m^{\prime };\ t)=\int_{0}^{1}d\beta \frac{\sin (m\tau _{\beta }t/2)}
{\sin (m\tau _{\beta }/2)}\frac{\sin (m^{\prime }\tau _{\beta }t/2)}{\sin
(m^{\prime }\tau _{\beta }/2)}C_{\beta }(m-m^{\prime })e^{i(m^{\prime
}-m)(t+1)\tau _{\beta }/2}.  \label{Bmt}
\end{equation}
We show in the Appendix that the asymptotic behavior of the quantity (\ref
{Bmt}) for $t\rightarrow \infty $ is given by 
\begin{equation}
B(m,\ m^{\prime };\ t)\sim \frac{t}{2mm^{\prime }l_{0}}\left( |m|+|m^{\prime
}|-|m-m^{\prime }|\right) \sum_{r=0}^{g(m,m^{\prime })l_{0}-1}C_{\beta
_{r,g}}(m-m^{\prime }),  \label{Bmt1}
\end{equation}
where $g=g(m,\ m^{\prime })$ is the greatest common factor of $(|m|,\
|m^{\prime }|)$ and $\beta _{r,g}=r/(gl_{0})-1/2$ ${\rm mod}(1)$, $r=0,\dots
,\ gl_{0}-1$. Rels. (\ref{ake1}) and (\ref{Bmt1}) imply a diffusive behavior
of $\left\langle E\right\rangle _{t}$ for large $t$, $\left\langle
E\right\rangle _{t}\sim Dt$. The diffusion coefficient $D$ can be expressed,
after some algebra, as the sum of two terms: 
\begin{eqnarray}
D &=&D_{{\rm I}}+D_{{\rm II}}=\frac{k^{2}}{l_{0}}\sum_{m>0}m|V_{m}|^{2}
\sum_{r=0}^{ml_{0}-1}C_{\beta _{r,m}}(0)  \nonumber \\
&&+\frac{2k^{2}}{l_{0}}\sum_{m=1}^{\infty }\sum_{m^{\prime }=m+1}^{\infty }m
{\rm Re}\left( V_{m}V_{m^{\prime }}^{\ast }\right) \sum_{r=0}^{g(m,m^{\prime
})l_{0}-1}{\rm Re}\left[ C_{\beta _{r,g}}(m-m^{\prime })\right] .  \label{D}
\end{eqnarray}
For a potential containing harmonics $V_{m}$ of sufficiently high order $m$,
the term $D_{{\rm II}}$ [on the second line of Eq. (\ref{D})] can generally
lead to a sensitive dependence of $D$ on the correlations $C_{\beta }(m)$,
which reflect the profile of the initial wave packet in momentum space by
Eq. (\ref{cbm}). This term vanishes in some cases, e.g., for $V(\theta
)=\cos (\theta )$ and/or a uniform probability distribution $\left| \psi
_{\beta }(\theta )\right| ^{2}$ in Eq. (\ref{pbc}). One is then left with
only the first term ($D_{{\rm I}}$), which is completely analogous to the
diffusion coefficient (\ref{DIM}) for the incoherent mixture; this is
because $C_{\beta }(0)=\sum_{n}\left| \widetilde{\Psi }(n+\beta )\right| ^{2}
$ from Eq. (\ref{cbm}) and this is analogous to $f_{0}(\beta
)=\sum_{n}f(n+\beta )$.

\begin{center}
{\bf V. MOMENTUM PROBABILITY DISTRIBUTION}
\end{center}

The momentum probability distribution (MPD) for a KP wave packet is given
by $P(p,\ t)=\left| \widetilde{\Psi }_{t}(p)\right| ^{2}$, where $\widetilde
{\Psi }_{t}(p)$ is the momentum representation of the wave packet at time 
$t$. At fixed $\beta $, $P(n+\beta ,\ t)$ is the angular-momentum ($n$)
distribution for a $\beta $-KR. Under QR conditions and for resonant $\beta$,
the motion of the $\beta$-KR is ballistic in $n$, so that the width
$\Delta n(t)$ of $P(n+\beta ,\ t)$ increases much faster than that for a
nonresonant $\beta $ [for most values of $\beta $, $\Delta n(t)$ is expected
to be essentially bounded due to dynamical localization]. Thus, if $p$ and $t
$ are sufficiently large, $P(p,\ t)$ is almost zero except of narrow peaks
around $p=n+\beta _{r,g}$, for all resonant values $\beta _{r,g}$ of $\beta$.
The diffusion coefficient (\ref{D}) is the average of $P(p,\ t)p^{2}/(2t)$
over $p$ in the limit of $t\rightarrow \infty $; the only nonzero
contributions to this average come from the resonant peaks. The contribution
of all the peaks with fixed $\beta _{r,g}=\beta _{r^{\prime },g^{\prime }}$
is precisely the contribution of all the terms with $\beta _{r,m}=\beta
_{r^{\prime },g^{\prime }}$ and/or $\beta _{r,g}=\beta _{r^{\prime
},g^{\prime }}$ in formula (\ref{D}).\newline

We now study the MPD for $\tau =2\pi l_{0}$ and $V(x)=\cos (x)+\gamma \cos
(2x)$ (see also end of Sec. III). From Rel. (\ref{phb}), $\widetilde{\Psi }
_{t}(n+\beta )$ are just the Fourier coefficients of $\sqrt{2\pi }\psi
_{\beta ,t}(\theta )$ and, for $\tau =2\pi l_{0}$, $\psi _{\beta ,t}(\theta )
$ can be calculated using Rel. (\ref{pbt}). We assume an initial wave packet 
$\widetilde{\Psi }(p)$ satisfying $\widetilde{\Psi }(p)=1$ for $0\leq p<1$
and $\widetilde{\Psi }(p)=0$ otherwise. This corresponds, by Rel. (\ref{phb}),
to a uniform $\psi _{\beta }(\theta )$, $\psi _{\beta }(\theta )=(2\pi
)^{-1/2}$ for all $\beta $. Rel. (\ref{pbt}) then implies that $\sqrt{2\pi }
\psi _{\beta ,t}(\theta )=\exp \left[ -ik\bar{V}_{\beta ,t}(\theta )\right] $.
Since only terms with $|m|=1,\ 2$ appear in the sum (\ref{Vbt}), the
Fourier coefficients of $\exp \left[ -ik\bar{V}_{\beta ,t}(\theta )\right] $
can be easily expressed, essentially, as a convolution of ordinary Bessel
functions $J_{n}(\cdot )$ in the index $n$. We finally obtain the exact
expression 
\begin{equation}
P(p=n+\beta ,\ t)=\left| \sum_{m=-\infty }^{\infty }i^{m}J_{n-2m}\left[
k_{1}(\beta ,\ t)\right] J_{m}\left[ k_{2}(\beta ,\ t)\gamma \right] \right|
^{2},  \label{Pt}
\end{equation}
where $k_{j}(\beta ,\ t)=k\sin \left( j\tau _{\beta }t/2\right) /\sin \left(
j\tau _{\beta }/2\right) $, $j=1,\ 2$, and $\tau _{\beta }=\pi l_{0}(2\beta
+1)$. Let us consider some behaviors of (\ref{Pt}) for $\beta =\beta _{r,g}$,
where $\beta _{r,g}$ ($g=1,\ 2$) are the resonant values of $\beta $
determined in Sec. III: (a) For $\beta _{r,1}=r/l_{0}-1/2$ ${\rm mod}(1)$, 
$r=0,\ 1,\dots ,\ l_{0}-1$, one has $|k_{1}(\beta _{r,1},\ t)|=|k_{2}(\beta
_{r,1},\ t)|=kt$. Now, the Bessel function $J_{m}(x)$ is relatively small
for $|m|>|x|$ and, for $|x|<1$, $J_{0}(x)=O(1)$ \cite{gr}. We then see that
on the time scale $t\lesssim T_{\gamma }=|k\gamma |^{-1}$ one can
approximate (\ref{Pt}) at $p=n+\beta _{r,1}$ by $P(p,\ t)\approx
J_{n}^{2}(kt)$. Thus, the most prominent $g=1$ peaks of the MPD for
$t\lesssim T_{\gamma }$ are those with $|n|\lesssim |kt|$. As $\gamma
\rightarrow 0$ ($T_{\gamma }\rightarrow \infty $), the expression (\ref{Pt})
reduces exactly to $P(p,\ t)=J_{n}^{2}\left[ k_{1}(\beta ,\ t)\right] $ (see
note \cite{note2}). (b) For $\beta _{r,2}=r/(2l_{0})-1/2$ ${\rm mod}(1)$,
$r=1,\ 3,\dots ,\ 2l_{0}-1$, one has $|k_{1}(\beta _{r,2},\ t)|=k$ or $0$ for 
$t$ odd or even, respectively, and $|k_{2}(\beta _{r,2},\ t)|=kt$. Then, if
$|k|\lesssim 1$, $J_{n-2m}\left[ k_{1}(\beta _{r,2},\ t)\right] $ is
relatively small for $|n-2m|>1$ and one can approximate (\ref{Pt}) at
$p=n+\beta _{r,2}$ by $P(p,\ t)\approx J_{[n/2]}^{2}(k\gamma t)$, where 
$[n/2]$ is the integer part of $n/2$. The new ($g=2$) peaks thus start to 
emerge in a significant way when $t>|k\gamma |^{-1}$ and, for $|n|\lesssim
2|k\gamma t|$, their magnitude should be comparable to that of the $g=1$
ones. All these behaviors are illustrated in Fig. 1 for $\tau =2\pi $ 
($l_{0}=1$), $t=100$, $k=0.1$, $\gamma =0$ [Fig. 1(a)] and $\gamma =0.2$ [Fig.
1(b), with $2|k\gamma t|=4$]. We have checked numerically for many values of 
$(t,\ k,\ \gamma )$ that the MPD is robust under sufficiently small
perturbations of $\tau $, $\tau =2\pi +\epsilon $, at least within the
domains of $p$ where the principal peaks above are found. As an example,
compare Fig. 1(c) with Fig. 1(b).

\begin{center}
{\bf VI. SUMMARY AND CONCLUSIONS}
\end{center}

In conclusion, the results in this paper should provide new insights into
the nature of the spectra and quantum dynamics of the KP under general QR
conditions and for arbitrary potentials. As a direct consequence of
translational invariance, any rational value of the Bloch quasimomentum 
$\beta $ may be resonant (exhibits QR). At fixed $\tau /(2\pi )=l_{0}/q_{0}$ 
($l_{0}$ and $q_{0}$ are coprime integers), a rational $\beta $ is characterized 
by an integer pair $(r,\ g)$ which determines $\beta $ through Eq. (\ref{br}), 
with $l=gl_{0}$ and $q=gq_{0}$ assuming their minimal values. The QE spectrum 
consists of $q=gq_{0}$ bands which should be, typically, not all flat, 
implying QR. By slightly varying $\beta $ on the rationals, the corresponding 
value of $g$ changes erratically, leading to similar changes in the QE spectrum. 
The important case of the main QRs ($q_{0}=1$) is precisely described by the linear 
KR and is thus exactly solvable. In this case, all rational values of $\beta $ are 
indeed resonant for generic potentials containing all the harmonics. In general, 
the resonant values of $\beta $ correspond to peaks in the momentum probability 
distribution (see Fig. 1) and appear explicitly in the formulas (\ref{DIM}) and 
(\ref{D}) for the diffusion coefficients. Formula (\ref{D}) implies a new 
phenomenon for multi-harmonic potentials [thus excluding the standard case of 
$V(x)=\cos (x)$]: A sensitive dependence of the diffusion coefficient on the 
specific localization features of the initial wave packet in momentum space. 
Assuming the robustness of our results under small variations of $\tau $, 
which was numerically verified for a two-harmonic potential, it may be possible 
to observe this phenomenon in experimental realizations of the system. While
high-order ($q_{0}>1$) QRs appear to be presently beyond experimental
observation, an interesting question is whether cases of such QRs are
exactly solvable, at least to some extent, for the $\beta $-dependent QE
spectra and quantum dynamics. We do not have yet a definite answer to this
question. We hope that we shall be able to make some progress in this
direction in future works.

\begin{center}
{\bf ACKNOWLEDGMENTS}
\end{center}

This work was partially supported by the Israel Science Foundation (Grant
No. 118/05). DLD acknowledges partial support from the Russian Ministry of
Education and Science and the US Civilian Research and Development
Foundation (CRDF BRHE Program, Grants Nos. VZ-0-010 and Y2-P-10-01).

\begin{center}
{\bf APPENDIX}
\end{center}

We derive here the asymptotic behavior (\ref{Bmt1}). The dominant
contributions to the integral (\ref{Bmt}) come from small $\beta $-intervals
around the zeros of the denominator of the integrand. We show below that the
behavior (\ref{Bmt1}) is completely due to the simultaneous zeros of $\sin
(m\tau _{\beta }/2)$ and $\sin (m^{\prime }\tau _{\beta }/2)$ ($m,\
m^{\prime }\neq 0$), $\tau _{\beta }=\pi l_{0}(2\beta +1)$. The zeros of,
say, $\sin (m\tau _{\beta }/2)$ are $\beta =\beta _{r,m}=r/(|m|l_{0})-1/2$
${\rm mod}(1)$, $r=0,\dots ,\ |m|l_{0}-1$. Writing $m=gm_{0}$ and $m^{\prime
}=gm_{0}^{\prime }$, where $m_{0}$ and $m_{0}^{\prime }$ are coprime
integers and $g=g(m,\ m^{\prime })$ is the greatest common factor of $(|m|,\
|m^{\prime }|$ $)$, it is easy to see that there are precisely $gl_{0}$
simultaneous zeros, given by $\beta =\beta _{r,g}=r/(gl_{0})-1/2$ 
${\rm mod}(1)$, $r=0,\dots ,\ gl_{0}-1$.\newline

Thus, let $\bar{\beta}=\beta _{r,m}$ be a zero of $\sin (m\tau _{\beta }/2)$
which is not a zero of $\sin (m^{\prime }\tau _{\beta }/2)$, $|m^{\prime
}|\neq |m|$, and consider a $\beta $-interval $I_{\epsilon }=[\bar{\beta}
-\epsilon ,\ \bar{\beta}+\epsilon ]$ sufficiently small that no zero of
$\sin (m^{\prime }\tau _{\beta }/2)$ lies within it. We show that the
contribution of $I_{\epsilon }$ to (\ref{Bmt}) is finite in the limit of
$t\rightarrow \infty $. Let $|m|\pi l_{0}\epsilon \ll 1$, so that $\sin
(m\tau _{\beta }/2)\approx (-1)^{r}m\pi l_{0}(\beta -\bar{\beta})$ in
$I_{\epsilon }$. We assume that the correlation function (\ref{cbm}) can be
expanded as a Taylor series around $\beta =\bar{\beta}$: $C_{\beta }(m)=
C_{\bar{\beta}}(m)+C_{\bar{\beta}}^{\prime }(m)(\beta -\bar{\beta})+\cdots $.
The dominant contribution of $I_{\epsilon }$ to (\ref{Bmt}) is then
approximately given by 
\begin{equation}
B(I_{\epsilon })\approx \frac{(-1)^{r}C_{\bar{\beta}}(m-m^{\prime })}{m\pi
l_{0}}\int_{\bar{\beta}-\epsilon }^{\bar{\beta}+\epsilon }d\beta \frac{\sin
(m\tau _{\beta }t/2)}{\beta -\bar{\beta}}\frac{\sin (m^{\prime }\tau _{\beta
}t/2)}{\sin (m^{\prime }\tau _{\beta }/2)}e^{i(m^{\prime }-m)(t+1)\tau
_{\beta }/2}\text{.}  \label{BIe}
\end{equation}
We introduce the variable $z=\pi l_{0}(\beta -\bar{\beta})t$ and define
$\bar{z}=\pi l_{0}(\bar{\beta}+1/2)t$. For $t\gg \epsilon ^{-1}$, one can
see that $B(I_{\epsilon })$ in Eq. (\ref{BIe}) is well approximated by 
\begin{equation}
B(I_{\epsilon })\approx \frac{(-1)^{r(t+1)}C_{\bar{\beta}}(m-m^{\prime })}
{m\pi l_{0}\sin (m^{\prime }\tau _{\bar{\beta}}/2)}\int_{-\infty }^{\infty }
\frac{dz}{z}\sin (mz)\sin \left[ m^{\prime }(z+\bar{z})\right]
e^{i(m^{\prime }-m)(z+\bar{z})}.  \label{BIez}
\end{equation}
The integral in Eq. (\ref{BIez}) can be calculated exactly using simple
trigonometry and formulas (3.741.2) and (3.763.2) in Ref. \cite{gr}; its
value is finite for all $\bar{z}$ (or $t$).\newline

Let us therefore consider a simultaneous zero $\beta =\beta _{r,g}$ of $\sin
(m\tau _{\beta }/2)$ and $\sin (m^{\prime }\tau _{\beta }/2)$, denoting it
again by $\bar{\beta}$. The interval $I_{\epsilon }$ is defined as above
with $\pi l_{0}\epsilon \ll \min \left( |m|^{-1},\ |m^{\prime
}|^{-1}\right) $. The contribution of $I_{\epsilon }$ to (\ref{Bmt}) is,
approximately, 
\begin{equation}
B(I_{\epsilon })\approx \frac{(-1)^{r(m_{0}+m_{0}^{\prime })}}{mm^{\prime
}\pi ^{2}l_{0}^{2}}\int_{\bar{\beta}-\epsilon }^{\bar{\beta}+\epsilon
}d\beta \frac{\sin (m\tau _{\beta }t/2)\sin (m^{\prime }\tau _{\beta }t/2)}
{\left( \beta -\bar{\beta}\right) ^{2}}C_{\beta }(m-m^{\prime
})e^{i(m^{\prime }-m)(t+1)\tau _{\beta }/2},  \label{BIes}
\end{equation}
where $m_{0}$ and $m_{0}^{\prime }$ are the integers defined above.
Expanding again $C_{\beta }(m-m^{\prime })$ as a Taylor series around $\beta
=\bar{\beta}$, we see that the first-order term $C_{\bar{\beta}}^{\prime
}(m-m^{\prime })(\beta -\bar{\beta})$ in this expansion gives a contribution
to (\ref{BIes}) which is similar to (\ref{BIe}) and is thus finite for all 
$t$. The contributions of higher-order terms are, obviously, also finite for
all $t$. After simple algebra, one can easily verify that the contribution
$B^{(0)}(I_{\epsilon })$ of the zero-order term $C_{\bar{\beta}}(m-m^{\prime
})$ is well approximated, for $t\gg \epsilon ^{-1}$, by 
\begin{equation}
B^{(0)}(I_{\epsilon })\approx \frac{C_{\bar{\beta}}(m-m^{\prime })t}
{mm^{\prime }\pi l_{0}}\int_{-\infty }^{\infty }dz\,z^{-2}\sin (mz)\sin
(m^{\prime }z)\cos \left[ (m-m^{\prime })z\right] ,  \label{BIesz}
\end{equation}
where the variable $z$ was defined above. By expressing the product $\sin
(m^{\prime }z)\cos \left[ (m-m^{\prime })z\right] $ as a sum, the integral
in Eq. (\ref{BIesz}) can be calculated exactly using formula (3.741.3) in
Ref. \cite{gr}. The final result, with $\bar{\beta}=\beta _{r,g}$, is 
\begin{equation}
B^{(0)}(I_{\epsilon })\approx \frac{t}{2mm^{\prime }l_{0}}\left(
|m|+|m^{\prime }|-|m-m^{\prime }|\right) C_{\beta _{r,g}}(m-m^{\prime }).
\label{BIesze}
\end{equation}
Summing (\ref{BIesze}) over all the $gl_{0}$ simultaneous zeros $\beta
=\beta _{r,g}$ of $\sin (m\tau _{\beta }/2)$ and $\sin (m^{\prime }\tau
_{\beta }/2)$, we obtain the asymptotic behavior (\ref{Bmt1}).

\figure{FIG. 1. Momentum probability distribution (\ref{Pt}) for $t=100$, $k=0.1$, 
and (a) $\tau =2\pi$, $\gamma =0$; (b) $\tau =2\pi$, $\gamma =0.2$; (c) $\tau =
2\pi + \epsilon$ [$\epsilon =\pi (\sqrt{5}-1)/1200 \approx 0.0032$], $\gamma =0.2$. 
All the resonant peaks for $\gamma =0$ [in (a)] are at $p=n+1/2$. For $\gamma =0.2$ 
[in (b) and (c)], new peaks appear at integer $p=n$.\label{f1}}


\begin{references}
\bibitem{note}  In Ref. \cite{lkr2}, the potential (\ref{V}) was assumed to
be even (real $V_{m}=V_{-m}$). We present here the straightforward extension
of some results in Ref. \cite{lkr2} to the most general $V(\theta )$.
\bibitem{note1}  If $\tau /(2\pi )$ is rational but noninteger ($q_{0}>1$),
resonant values of $\beta =\beta _{r,g}$ with $g>1$ may exist already for
$V(\theta )=\cos (\theta )$. For example, in the case of $\tau /(2\pi )=1/2$
and $V(\theta )=\cos (\theta )$, we found numerically that for $\beta =1/2$
[minimal $g=2$, $r=1$, and $l=gl_{0}=2$ in Eq. (\ref{br})] all the
$q=gq_{0}=4$ QE bands have nonzero width, implying QR.
\bibitem{note2}  Due to the uniformity of $\psi _{\beta }(\theta )$, Eq.
(\ref{Pt}) gives also the angular-momentum distribution for a $\beta $-KR
initially in the angular-momentum state $n=0$. Then, the $\gamma =0$
expression $P(n+\beta ,\ t)=J_{n}^{2}\left[ k_{1}(\beta ,\ t)\right] $ is 
just the known \cite{kp2} exact result for this distribution in the case of 
$V(\theta )=\cos (\theta )$.
\end{references}
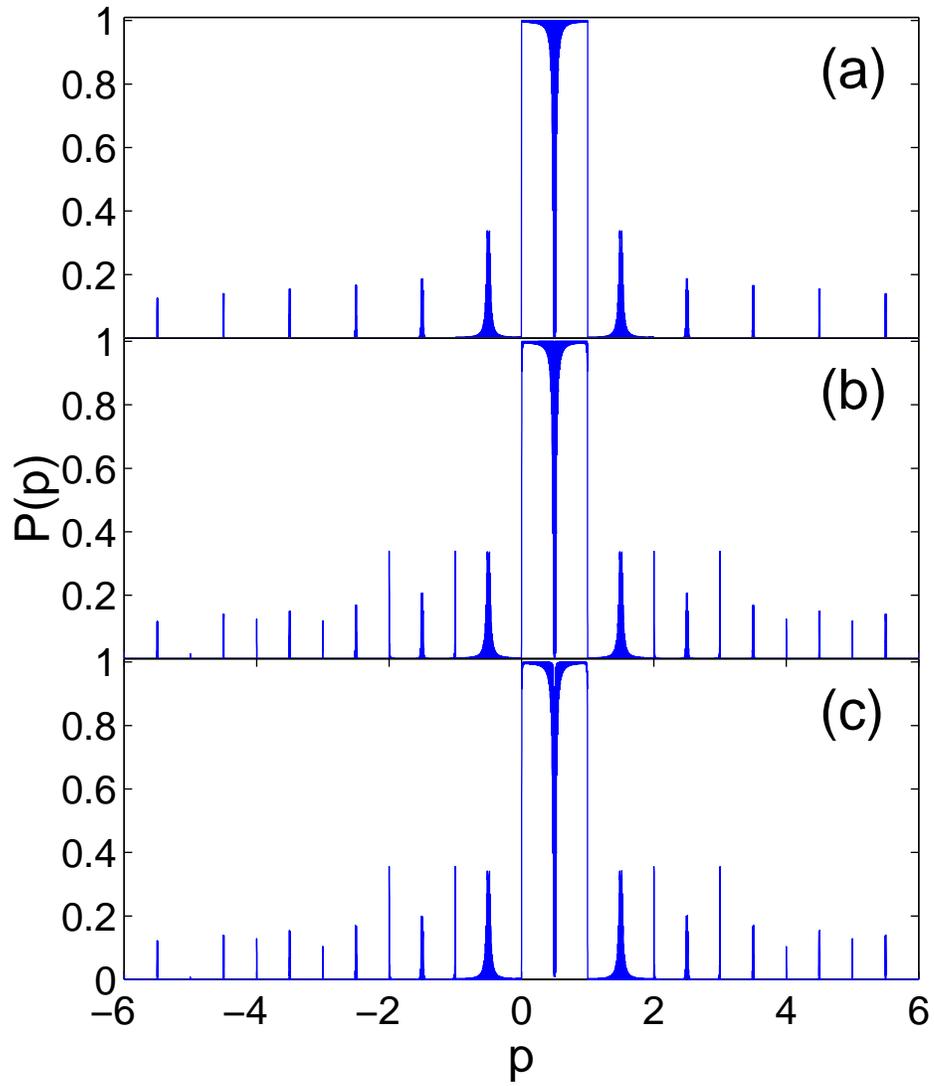
\end{document}